\documentclass[graybox]{svmult}

\usepackage{natbib}   

\usepackage{mathptmx}       
\usepackage{helvet}         
\usepackage{courier}        
\usepackage{type1cm}        
\usepackage{makeidx}         
\usepackage{graphicx}        
\usepackage{multicol}        
\usepackage[bottom]{footmisc}

\usepackage{url}        

\usepackage{xspace}
\usepackage[table]{xcolor}
\usepackage{amsmath}
\usepackage{amssymb}
\graphicspath{{figures/}}

\newcommand\simgreater{\,\lower0.7ex\hbox{$\stackrel{>}{\sim}$}\,}
\newcommand\simless{\,\lower0.7ex\hbox{$\stackrel{<}{\sim}$}\,}

\newcommand{\gcc}{\ \mathrm{g\ cm^{-3} }}

\newcommand{\microm}{\ \mathrm{\mu m}}

\makeindex             

\begin{document}


\title*{Astrophysical Validation}

\author{A. C. Calder and D. M. Townsley}

\institute{A. C. Calder \at Stony Brook University, Stony Brook, NY
11794-3800, USA \email{alan.calder@stonybrook.edu}
\and D. M. Townsley \at University of Alabama, Tuscaloosa, AL 35487-0324, USA \email{Dean.M.Townsley@ua.edu}}

\maketitle

\abstract{
We present examples of validating components of an astrophysical
simulation code. Problems of stellar astrophysics are multi-dimensional
and involve physics acting on large ranges of length and time scales that
are impossible to include in macroscopic models on present computational
resources. Simulating these events thus necessitates the development
of sub-grid-scale models and the capability to post-process simulations
with higher-fidelity methods. We present an overview of the problem of
validating astrophysical models and simulations illustrated with two
examples. First, we present a study aimed at validating hydrodynamics
with high energy density laboratory experiments probing shocks and
fluid instabilities. Second, we present an effort at validating code
modules for use in both macroscopic simulations of astrophysical events
and for post processing Lagrangian tracer particles to calculate
detailed abundances from thermonuclear reactions occurring during 
an event.}

\section{Introduction}

Verification and validation (V\&V) of models and simulations of
astrophysical phenomena present challenges because the problem
of studying these phenomena is largely one of indirectly observing
multi-scale, multi-physics events. Other aspects of astrophysics
also contribute challenges. The enormous length scales 
of astrophysical objects and vast distances to most astrophysical events 
preclude ready experimental access, limiting
the availability of validation data.  As with a great many applications,
models suffer from epistemic uncertainty in the underlying basic
physics (e.g.\ turbulence, fluid instabilities, and nuclear reaction rates),
which is difficult to
control and assess in simulations incorporating multiple interacting
physical processes.  The large range of length and time scales in many
astrophysical problems frequently necessitates capturing sub-grid-scale
physics within simulations, relevant examples being thermonuclear flames
and turbulent combustion.  The requirement of the development of
sub-grid-scale models for these physical processes obviously introduces
an additional level of complexity to V\&V. Finally, the magnitude of the
requisite computations for astrophysical events means that even with
sub-grid-scale models, simulations may only capture the bulk effect 
of the underlying physics and some properties such as detailed compositions 
must be obtained by post-processing the simulation results with augmenting, 
higher-fidelity routines.

Even with these issues, V\&V are vital parts of computational astrophysics 
as with any research domain. 
We present two studies aimed at validating components of Flash, a freely
available, parallel, adaptive mesh simulation code used for modeling 
astrophysical phenomena and other applications. 
We first present a study of validating the hydrodynamics routines in 
Flash with experiments designed to replicate the high energy
density environments of astrophysics and probe the
underlying physics. The investigation formally addresses the issues of 
concern in validating hydrodynamics and serves as a well-controlled case study. 
The second study we present addresses physics that is difficult
to include in whole-star simulations, due to limits in computing power, but that 
can be incorporated with approximate models and also calculated by 
post-processing simulation results. The problem is thermonuclear
combustion and describing the overall reactions while including
minimal nuclear species, and this work addresses the issue 
of comparing
prohibitively expensive detailed models and simpler
models that allow three-dimensional simulations.

As we will describe below, the challenges to astrophysical validation
made parts of our study incomplete. The effort, however, was rewarding 
and very much worth the investment. Verification tests quantified
the accuracy of code modules for problems with an analytic or accepted
result, and the regular application of these tests 
serves for regression testing as the code is developed. Validation tests,
though incomplete, demonstrated reasonable agreement between experiment and 
simulation for the case of the hydrodynamics study. Comparison between
models of increasing sophistication allowed us to quantify the trade-off
between fidelity of the method and expense.
These studies all led to a deeper understanding of the
underlying physics, and while we cannot say the modules and code were 
completely ``validated,"
the process greatly increased our confidence in the results.  

\section{Approach to Verification and Validation}
\label{sec:vvapproach}

Our methods for V\&V largely follow accepted practices from
the fluid dynamics community
\citep[see Ch.\ 26 by Roache in this volume]{1998aiaa,roache1998verification,roache98,oberkampf2010verification}.
We adopt the following definitions (based on definitions from
the American Institute of Aeronautics and Astronautics 
\citep{1998aiaa}). 
\begin{quotation}
\begin{trivlist}
\item Model: A representation of a physical system or process intended to
enhance our ability to understand, predict, or control its behavior.
\item Simulation: The exercise or use of a model. (That is, a model is used
in a simulation).
\item Verification: The process of determining that a model implementation
accurately represents the developer's conceptual description of the
model and the solution of the model.
\item Validation: The process of determining the degree to which a
model is an accurate representation of the real world from the perspective
of the intended uses of the model.
\item Uncertainty: A potential deficiency in any phase or activity of the modeling
process that is due either to a lack of knowledge (epistemic uncertainty or
incertitude) or due to variability or inherent randomness (aleatory uncertainty).
\item Error: A recognizable deficiency in any phase or activity of modeling that is
not due to lack of knowledge.
\item Prediction: Use of a model to foretell the state of a physical system under
conditions for which the model has not been validated.
\item Calibration: The process of adjusting numerical or physical modeling parameters
in the computational model for the purpose of improving agreement with experimental
data.
\end{trivlist}
\end{quotation}
Our definition of uncertainty differs from the original definition of the
AIAA in that we expand the definition of uncertainty to also include aleatory 
uncertainty~\citep[see][and references therein]{uqproceedings2018,hoffmanetall2018}.

Another perspective comes from \citet{roache98}, who offers a concise, 
albeit informal, 
summary of V\&V terminology:
\begin{quotation}
First and foremost, we must repeat the essential distinction between
Code Verification and Validation.
Following \citet{boehm81} and \citet{blottner90}, we adopt the
succinct description of ``Verification" as ``solving the equations right",
and ``Validation" as ``solving the right equations". The code author defines
precisely what partial differential equations are being solved, and convincingly
demonstrates that they are solved correctly, i.e. usually with some order of accuracy,
and always consistently, so that as some measure of discretization 
(e.g.\ the mesh increments)
$\Delta \rightarrow 0$, the code produces a solution to the continuum equations; this
is Verification. Whether or not those equations and that solution bear any relation
to a physical problem of interest to the code user is the subject of Validation.
\end{quotation}
Roache also notes that a ``code" cannot be validated, but only a calculation or 
range of calculations can be validated. Roache also makes a distinction between 
verifying a code and verifying a calculation, noting that ``use of a verified 
code is not enough." We also adhere to this explication of V\&V terminology 
and
note that following Roache, validation can be described as probing the range 
of validity of a code or model~\citep{calder.fryxell.ea:on}.

Our approach to verification consists of testing simulation results against 
analytic or benchmarked solutions and quantifying the error. The comparisons 
typically consist of simulations performed at increasing spatial and/or temporal
resolutions to confirm convergence of the simulation to the correct answer. 
Details of these tests have appeared in the literature,
and many of the tests are incorporated into automated regression testing of Flash
\citep{calder.fryxell.ea:on, weirsetal2005a, weirsetal2005b, dwarkadasetal2005, hearnetal2007,
dubeyetal2009, dubeyetal2015}. 

We validate by performing similar tests against data from experiments
designed to replicate astrophysical environments.
We take a hierarchical approach to validation, beginning
by isolating the basic underlying physics and testing
how well simulations capture it. We then devise tests of 
aggregate problems that capture the expected behavior of the 
astrophysical events. In the case of sub-grid
models or post-processed results, we simulate simple problems with these
models and compare against either actual validation data or direct numerical
simulations.  As with verification, we perform convergence 
tests, though as we describe below the process 
of demonstrating convergence is difficult for some fluid dynamics problems.

Another aspect of our testing concerns quantifying
error on the adaptive simulation mesh (described below).
Our approach is to test solutions on the finest simulation mesh
against data or a solution, but the methodology for quantitatively comparing
the solution at the different resolutions of an adaptive mesh is incomplete 
\citep{li2010,crash,shuetal2011,liwood2017}. We typically check for 
consistency between simulations on an adaptive mesh and simulations
of the same problem on a fully-refined mesh while quantifying the
accuracy of the solution on the fully-refined mesh \citep{calder.fryxell.ea:on}. 
Also, in addition to problems characterizing solutions on an
adaptive mesh, just simulating fluids at the extreme Reynolds numbers of 
astrophysics 
on adaptive meshes presents challenges~\citep{kritsuknormanpadoan2006,mitran2009}. 
We describe the difficulties of simulating extreme Reynolds number flow in
the discussion of our hydrodynamics method below.

We close discussion of our approach to V\&V with a general note on the role 
of validation in astrophysics.
Because of the literally astronomical distances to astrophysical
events and extreme conditions involved, 
experimentally accessing astrophysical phenomena or even just replicating 
the environments of astrophysics is difficult. Thus one cannot readily 
perform validation experiments, which typically leads to an incomplete 
process of validation. Simulations of astrophysical events are therefore
generally in the realm of prediction, that is, foretelling the state of a 
physical system under conditions for which the model has not been validated. 
Despite this, the process of V\&V in astrophysics serves to
build confidence in these predictions even if one cannot conclude that
simulations or codes are ``validated."

\section{Simulation Instruments}

Our principal simulation instrument is the Flash code, which we use for 
simulating astrophysical events. Fundamentally, Flash
simulates problems of fluid dynamics and consists of solvers for hydrodynamics
and the additional physics of astrophysical events (described below). With
Flash, we construct the numerical implementation of our conceptual model
of the astrophysical event, and the act of simulating is the exercise of the 
model. We note that the exercise of a model is far more than just solving
a set of differential equations. Multi-physics applications
like astrophysics combine multiple solvers, each of which may rely on possibly
uncontrolled assumptions~\cite[See][for a thorough discussion]{winsberg2010science}.
For this reason, we take the hierarchical approach to validation of modules
in Flash mentioned above.

Our second instrument is a nucleosynthetic post-processing toolkit used in tandem with Flash.
In the case of supernovae, comparison to observations requires the calculation
of light curves (the intensity of light from the object as a function of time) and spectra.
However, the yield of a particular element, titanium for example, may be critical for accurate spectra, but mostly unimportant to the energy release.
Many elements fall into this category, so that the computation of the explosion is much less expensive when split into two stages.
The energy release and explosion is computed with a small number of species in Flash, and is followed by post-processing to obtain all important species.
The post-processing tools we present below apply state-of-the-art nuclear reaction networks to Lagrangian thermodynamic histories sampled from the Flash simulation.
The resulting abundances are used to calculate light curves and spectra~\citep[e.g.][]{milesetal2016}.

\subsection{The Flash Code}

The simulation instrument we use for
modeling astrophysics events is the Flash code, developed at the University
of Chicago \citep{Fryxetal00, calder.curtis.ea:high-performance,
dubeyetal2009,dubeyetal2013,dubeyetal2014}. Flash is a parallel, adaptive-mesh, 
hydrodynamics plus additional physics code originally designed for the 
compressible fluid flows associated with astrophysics. Flash incorporates
multiple hydrodynamics methods \citep{Fryxetal00,LEE2009952,LEE2013269,Lee2017230,leeetall2017}
coupled with modules for the requisite additional physics 
of the applications. In particular, Flash has undergone considerable development
for high energy density physics applications \citep{Tzeferacosetal2015}.

The hydrodynamics
modules solve the Euler 
equations of compressible hydrodynamics, shown here with gravitational
sources as would apply to a self-gravitating problem such as a star.
\begin{displaymath}
\frac{\partial \rho}{\partial t} +  {\bf \nabla} \cdot  
(\rho {\bf v}) = 0 
\end{displaymath}

\begin{displaymath}
\frac{\partial \rho {\bf v}}{\partial t} +  {\bf \nabla} \cdot
\left(\rho {\bf v}{\bf v}\right)  +  {\bf \nabla}P = \rho {\bf g}  
\end{displaymath}

\begin{displaymath}
\frac{\partial \rho E}{\partial t} 
+  {\bf \nabla} \cdot \left(\rho E + P\right) {\bf v} \: = \: 
\rho {\bf v} \cdot {\bf g} + S \; .
\end{displaymath}
Here $\rho$ is the mass density, ${\bf v}$ is the velocity, $P$ is
the pressure, $E$ is the internal energy of the gas, ${\bf g}$ is the 
gravitational acceleration, and $S$ represents any additional source. 
The system is closed by an equation of state of the form
\begin{displaymath}
P = P\left(\rho,E\right)
\end{displaymath}
and Flash offers choices for particular applications. 
Flash calculates the acceleration due to gravity
from the gravitational potential,
\begin{displaymath}
{\bf g} = - {\bf \nabla} \Phi \; .
\end{displaymath}
which is calculated by solving the 
Poisson equation for self-gravity
\begin{displaymath}
\nabla^2 \Phi\left({\bf r}\right) = 4\pi G \rho\left({\bf r}\right) \; .
\end{displaymath}
Here $\phi$ is the gravitational potential and $G$ is Newton's 
gravitational constant.
Flash also tracks different material species by
advecting mass scalars,
\begin{displaymath}
\frac{\partial X\rho}{\partial t} +  {\bf \nabla} \cdot  
(X \rho {\bf v}) = 0 \; ,
\end{displaymath}
where $X$ is the mass fraction of a given species~\citep{Fryxetal00}. 

Our first validation example
addressed the Flash hydrodynamics module (without gravity) for the case
of experiments involving fluid instabilities thought to occur
during one class of  stellar explosions known as a core collapse supernova~\citep{fryxellarnettmueller1991}.
The particular hydrodynamic module in Flash used for this study
is based on the PROMETHEUS code \citep{FryxMuelArne89} and evolves the Euler equations 
in one, two, or three dimensions using a modified version
of the Piecewise-Parabolic Method (PPM) \citep{Colella1984The-Piecewise-P}.
The implementation allows use of general equations of
state as is required for simulating stellar material \citep{colellaglaz85}, but
this capability was not used in the validation example.

PPM is a higher-order version of the method developed by Godunov~\citep{godunov59,godunov62},
a finite-volume conservation scheme that
solves the Riemann problem at the interfaces of the control
volumes to compute fluxes into each volume. The conserved fluid
quantities are treated as cell averages that are updated by the fluxes
at the interfaces. This treatment has the effect of introducing explicit
non-linearity into the difference equations and permits the calculation of
sharp shock fronts and contact discontinuities without introducing
significant non-physical oscillations into the flow. In addition,
PPM utilizes a dissipative shock capturing scheme to further stabilize
shocks and contact discontinuities, and is thus not directly solving the
Euler equations \citep{majda84,winsberg2010science}.

The adaptive mesh of Flash is block structured and is supported primarily
through the Paramesh Library~\citep{paramesh1,paramesh2}, though it is under the 
process of migrating to the AMReX library~\citep{amrex}. The view of AMReX from 
other units in the Flash code will remain similar to that of Paramesh, and in 
the near future the two packages will be available as alternative implementations 
of the Grid unit. Later, the support for Paramesh may be dropped if it becomes too 
inefficient on newer platforms. 

\subsection{The Post-processing Toolkit}

The nucleosynthetic post-processing toolkit uses the recorded Lagrangian history of fluid elements to compute the yield of nuclides (elements and their isotopes) produced in a stellar explosion
\citep{travaglioetal2004,townetal15}.
The Lagrangian thermodynamic history is determined by integrating the position of a conceptual microscopic fluid element by
\begin{displaymath}
{\bf r}(t) = {\bf r}_0 + \int_0^t {\bf v}({\bf r}, t')\,dt'\ ,
\end{displaymath}
where ${\bf r}_0$ is the initial position and ${\bf v}({\bf r},t)$ is the velocity field as computed by the hydrodynamic simulation.
This conceptual fluid element is often called a particle because it moves as a fluid-embedded particle would.
From the resulting ${\bf r}(t)$, it is possible to also record the thermodynamic state,
namely $T(t) = T({\bf r}(t))$ and $\rho(t)=\rho({\bf r}(t))$, the temperature and density, respectively.
Such recorded histories are often called tracks or trajectories
because they represent how the fluid element evolves in location and thermodynamic state space as a function of time.

Nucleosynthetic post-processing is performed in order to obtain the composition of material after it is processed by combustion and ejected.
Composition is parameterized by abundances of various species quantified as the fraction of a unit of mass that is in the form of a particular species.
For example, the fraction, by mass, that is in the form of $^{12}$C, may be written $X_{^{12}\rm C}$, and must be between 0 and 1.
The abundances are found in post-processing by integrating
\begin{displaymath}
\label{eq:abund_integral}
X_i(t) = X_{i,0} + \int_0^t \dot X_i( \rho(t'), T(t') )\, dt'\ ,
\end{displaymath}
where $\dot X_i(\rho,T)$ are determined by the density and temperature-dependent reaction rates for processes which involve species $i$.
Any given specie is typically involved in multiple reactions, forming a network that is used to evaluate each rate.
The end of the necessary integrations is typically well defined.
As the star expands $T$ and $\rho$ fall until most reactions will become very slow compared to the time being simulated, effectively freezing out.
Consideration of further evolution, typically radioactive decay, may be necessary depending on the usage of the resulting abundances.
These integrations are typically performed for a large number of tracks which sample the ejected material by a suitably distributed choice of their initial positions ${\bf r}_0$.

\subsection{Simulating Reactive Flow}

With both Flash and the post-processing toolkit, the goal
of simulations is to capture the evolution of stellar material during the
course of an astrophysical event. Because stars are essentially self-gravitating
gas, the interiors of stars are well described by the equations of fluid flow.
During an astrophysical event, thermonuclear reactions drive the evolution by
changing the composition and by releasing energy, which changes thermodynamic 
conditions like pressure and density. This combustion 
typically occurs in a relatively small region of space, e.g.\ a thin
flame, that may be difficult to resolve in simulations of the event.
The two validation examples we present address the two principal parts- fluid flow, including
shocks and fluid instabilities, and the evolution of the composition. 

For fluid dynamics problems, there are two fundamental classes of simulation
distinguished by whether or not the scales of the numerical grid can resolve 
viscous effects~\citep[][and references therein]{calder.fryxell.ea:on,winsberg2010science}. 
Simulations that can resolve viscous effects are said to be ``Direct Numerical
Simulations,'' while those that cannot and rely on a (possibly uncontrolled)
sub-grid-scale model for viscous effects are referred to as ``Large Eddy
Simulations.'' An eddy is a fluid current whose flow direction differs from 
that of the general flow, and the motion of the fluid is the net result of 
the movements of the eddies that compose it~\citep{eddy_britanica}.
Large eddy simulations do not resolve either the explicit viscosity of the fluid
or the contribution to the viscosity from eddies on unresolved scales~\citep
[][and references therein]{Fureby1996,ZHIYIN201511}. 

The issue of convergence of a solution for fluid flows is not as simple 
as it might seem. The enormous size of objects means that astrophysical 
regimes typically have Reynolds numbers far in excess of the Reynolds numbers 
of terrestrial flows,
which are themselves higher than can be 
readily captured in hydrodynamics simulations. Even when run on 
contemporary supercomputers, simulations cannot capture the
possibly $\ge 10^{8}$
Reynolds numbers of astrophysical flows making direct numerical
simulations impossible. Thus simulations of astrophysical
events are large eddy simulations that can either rely on 
sub-grid-scale models for turbulent flow or just allow the
intrinsic numerical
diffusion of the hydrodynamics method to set the limiting Reynolds 
number. This 
latter case, know as Implicit Large Eddy Simulation (ILES), is 
frequently applied and is the approach taken in the studies presented
here. In ILES,
changing the resolution changes the effective viscosity and hence the
Reynolds number, which changes the problem itself and leads to the
question of convergence of results with resolution. Considerable
study has gone into determining the validity of this approach
\citep{margolinrider2002,grinstein2007implicit, margolinshashkov2008,
margolin2014}. As our results show, large eddy simulations may not 
demonstrate convergence of a solution with resolution.

\section{Validation Examples}

As of this writing, Flash has had 20 years of development by generations
of scientists. Much of this effort has been subjected to rigorous V\&V
\citep{calder.fryxell.ea:on, cise_val_2004, weirsetal2005a, weirsetal2005b, dwarkadasetal2005, 
hearnetal2007, dubeyetal2009, dubeyetal2015,townetal15}. In this contribution, 
we present two examples of validating the Flash code and post-processing toolkit
for astrophysical applications.
The first example is from early work comparing simulations to laboratory 
experiments addressing
fluid instabilities in high energy density environments similar to the interiors of stars.
The second example is ongoing work on computing reaction products in 
three-dimensional simulations of type Ia supernovae.
This study includes comparison between methods for use in the simulations
of the events and for calculating detailed abundances from the
density and temperature histories of Lagrangian tracers. 

While this contribution describes two examples of the
V\&V efforts for the Flash code, we note that V\&V efforts continue
as the capabilities and applications of Flash evolve. A recent 
survey of software engineering practice in scientific computing includes Flash
as a case study and offers an independent perspective on the development
of Flash \citep{Storer:2017}.

\subsection{Overview of Flash Problems}
The Flash code was originally designed to investigate astrophysical
thermonuclear flashes, explosive events powered by thermonuclear
fusion. These events all involve a close binary star system with
matter being transferred (accreted) onto a compact star (either a
white dwarf or a neutron star) from a companion~\citep{rosneretal2000}.
The three flash problems originally addressed by Flash were type I x-ray
bursts~\citep{zingaleetal2001}, classical novae~\citep{alexakisetal2004},
and type Ia supernovae~\citep{plewacalderlamb2004,townsleyetal2007}.

X-ray bursts occur when a thermonuclear runaway occurs in a thin
$\sim10-100$ m layer of H- or He-rich fuel accreted onto the surface of
a neutron star. The radius of the underlying neutron star my be inferred
from observations and thereby allow constraints on the properties of
dense matter. Classical novae occur when a thermonuclear $\sim 10^4$
m thick layer of H-rich material similarly explodes. In this case,
material from the explosion is unbound and these events are thought to
synthesize some intermediate-mass elements found in the galaxy. Type Ia
supernovae are thought to occur when a pair of white dwarf star merge
and/or when a white dwarf accretes enough mass to ignite fusion in the
core. In this case,
enough energy is added to overcome the gravitational binding and the
star is completely disrupted, producing a bright explosion that
may be used as an indicator for cosmological distances.
(See references in above works for literature on each astrophysical topic,
and \citealt{calderetal2013} for an overview of ongoing investigation of Type Ia Supernovae.)

As mentioned above, these problems involve reactive flow, and in
all cases there is a vast difference between the length scale of the 
combustion front and the astrophysical object. Hence the need for sub-grid-scale
models. Fluid instabilities that may influence the burning rate are also of
particular importance~\citep{calderetal2007,zhangetal2007,townetal15}.
Accordingly, the  validation examples we present address problems of
combustion and fluid instabilities.

\subsection{Shocks and Fluid Instabilities}

The high energy density environments of intense lasers interacting 
with matter are similar to the interiors of stars, and experiments 
offer opportunities for a quantitative comparison between data and 
simulation not possible with observations of astrophysical phenomena.
The validation study we present was performed by a collaboration
between Flash developers and experimentalists working at the Omega 
laser at the University of Rochester~\citep{soures96,boehly95,
bradley98}.  The experiment chosen for the study
involved a shock propagating through a multi-layer target with 
layers of decreasing density and was designed to produce
hydrodynamic instabilities thought to arise during an astrophysical
event known as a core collapse
supernova explosion~\citep{arnett89,fryxellarnettmueller1991}.
While this type of supernova is not a thermonuclear flash problem, 
much of the constituent physics is the same, allowing this experiment 
to serve for validation. 
The decreasing-density configuration is subject to 
the Richtmyer-Meshkov instability
that occurs when a shock propagates though a material interface
with decreasing density~\citep{richtmyer60,meshkov69}. The configuration
is also subject to the Rayleigh-Taylor instability~\citep{taylor+50,chandra+81},
which develops after the passage of the shock and subsequently  dominates 
instability growth.

Interest in the problem of fluid instabilities during the
process of a core collapse supernova followed from the early observation
of radioactive elements from deep in the core of the star in SN 1987A
\citep{mulleretal1989}.  Stars with a mass of greater than 8-10 times
that of the Sun end their lives in a spectacular explosion known as
a core collapse supernova.  These events are among the most powerful
explosions in the cosmos, releasing energy of order $10^{53}$ erg at a
rate of $10^{45-46}$ watts, and are important for galactic
chemical evolution because they produce and disseminate heavy elements.
Core collapses supernovae also signal the birth of neutron stars and
black holes, which are the basic building blocks of other astrophysical
systems such as pulsars and x-ray binaries.

During their lifetimes, stars are powered by the thermonuclear fusion of
elements beginning with hydrogen fusing into helium. In a massive star,
fusion continues all the way to iron-group elements.  A core collapse
supernova occurs when the inert iron core can no longer support the
weight of the material above it and the core collapses, which releases
gravitational binding energy that is in part converted to the energy of
an expanding shock that ejects the outer layers of the star.  Just prior
to the explosion, the interior of the star has an onion-like structure,
with iron-group elements in the core, then layers of silicon, magnesium,
neon, oxygen, carbon, helium, and finally the outermost layer may be
hydrogen. When the supernova explosion occurs, the shock passes
through these layers of decreasing density. The early observation of a
core element suggests some sort mixing must have occurred during the
explosion, and, accordingly, motivated investigation into the effects
of fluid instabilities. The laboratory experiment was designed to probe
this scenario.

The experimental configuration consists of a strong
shock driven through a target with three layers of decreasing density.
The interface between the first two layers
is perturbed while the second interface is flat. An initially planar shock
created by the deposition of energy from the laser 
is perturbed as it crosses the first interface, which excites a Richtmyer-Meshkov
instability. As the perturbed shock propagates through the second interface,
the perturbation is imprinted on the interface. The material begins to flow,
leading to the growth of Rayleigh-Taylor instabilities. 
The three layers of the target are in a cylindrical 
shock tube composed of Be, with the density decreasing in the direction of
shock propagation. 
The materials were Cu, polyimide plastic, and
carbonized resorcinol formaldehyde (CRF) foam, with thicknesses of 85,
150, and 1500$\microm$ and densities 8.93, 1.41, and 0.1$\gcc$,
respectively. The shock tube delays the lateral decompression of the 
target, keeping the shock planar. The surface of the Cu layer was machined
with a sinusoidal ripple of wavelength $200\microm$ and amplitude $15\microm$
to perturb the shock as it passes this interface.
\begin{figure}[h]
   \includegraphics[angle=0,
      width=\columnwidth]{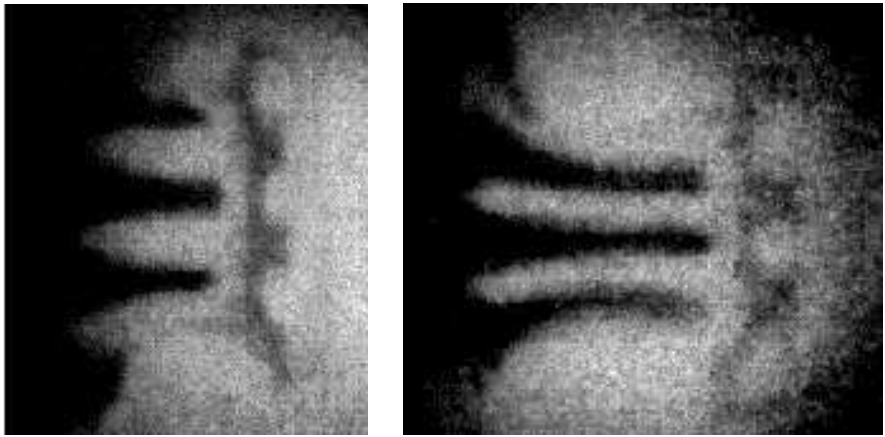}
   \caption{
Results of the three-layer target experiment. Shown are side-on X-ray radiographs at 
39.9 ns (left) and 66.0 ns (right). The long, dark ``fingers'' are
spikes of expanding Cu, and the horizontal band of opaque material to the right of 
the spikes of Cu is the brominated plastic tracer showing the imprinted
instability growth at the plastic-foam interface. From \citet{calder.fryxell.ea:on} \copyright\ AAS. Reproduced with permission.
      }
   \label{fig:shock_exp}
\end{figure}

The experiment was driven by 10 beams of the laser
with the target configured so that
the laser beams impinge a thin section of CH ablator to
prevent direct illumination and pre-heating of the target.
The experimental diagnostics were X-ray radiographs taken at different times during 
a ``shot." The Be shock tube, polymide plastic, and CRF 
foam are transparent to X-rays, while  the Cu layer is opaque to X-rays. Embedded within 
the polyimide layer was a tracer strip of brominated CH that is also opaque to 
X-rays. This tracer layer provided the diagnostic for polymide-foam interface, 
allowing visualization of the shock-imprinted structure. 

Figure \ref{fig:shock_exp} shows X-ray radiographs of the experiment at
two times, one relatively early at 39.9 ns (left) and one relatively
late at 66.0 ns (right). These images were from two different shots.
The long, dark ``fingers" are spikes of expanding
Cu, and the vertical band of opaque material to the right of the spikes of
Cu is the brominated plastic tracer, showing the imprinted instability growth
at the plastic-foam interface.  The radiographs illustrate the configuration at
early and late times in the evolution of the shocked target. 
The outer regions of the Cu and brominated strip show the effects of the 
shock tube, but the central part is largely immune to these effects. 

Making a quantitative comparison between the simulations and the experiments
and determining the uncertainty in the study required close collaboration
between experimentalists and theorists. This is an important point worth 
stressing. Without the contribution of both to interpreting and 
quantifying the experiments and simulations, there would have been 
little chance for a meaningful quantitative comparison. The data from the experiments 
are the radiographs, and finding a meaningful measurement for comparison to the 
simulation results required understanding the accuracy of the diagnostics
and sources of uncertainty in the experiment. The 
metric for comparison between simulation and experiment was chosen
as the length of the copper spikes, which are fairly obviously seen
in the radiograph, but which required a deep understanding of the experiments
to quantify. The paragraphs below summarize the sources of error and
uncertainty in the experiments and the reader is referred to the original
paper for complete details~\citep{calder.fryxell.ea:on}. A cautionary 
note concerning these details is
warranted, however. The intervening years between these experiments
and this writing have seen enormous progress in diagnosing high-energy-density
experiments and the experiments described here are not
the current state of the art~\citep{gamboa2012,stoecletal2012,gamboa2014a}.
\begin{figure}[t]
   \includegraphics[angle=0,
      width=\columnwidth]{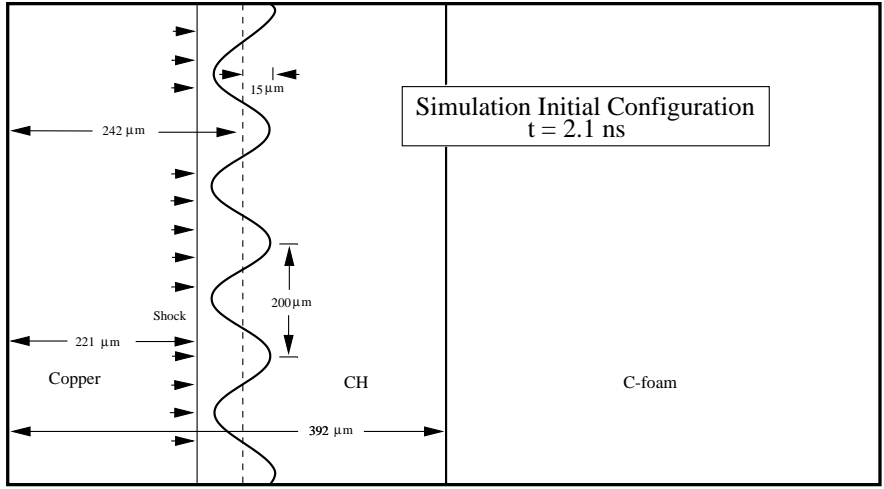}
   \caption{
Schematic of the three-layer target simulation initial conditions. Shown 
are the locations of the three materials, Cu, CH, and C, the shock, and
the details of the sinusoidal perturbation of the Cu-CH interface. The 
schematic is not to scale. From \citet{calder.fryxell.ea:on} \copyright\ 
	AAS. Reproduced with permission.
      }
   \label{fig:shock_init}
\end{figure}

The lengths of the Cu spikes in the experimental radiographs were determined 
by three methods.  The first was a straightforward visual inspection of the
images using a spatial reference grid located
just below the images of Figure \ref{fig:shock_exp}. The second used a contour
routine to better quantify the uncertainty in the location of the
edges of the spikes.  The third method was consistent
with the analysis of the simulations. A section in the
center of the images was vertically averaged to produce a single spatial
lineout of optical depth through the region occupied by the Cu and CH.
The same 5\% and 90\% threshold values were used to quantitatively determine
the extent of the Cu spikes.  Taking the average of all three methods,
values of 330 $\pm$ $25\microm$ and 554 $\pm$ $25\microm$ are obtained
at 39.9 and 66.0 ns, respectively.

Sources contributing to uncertainty in these experimental measurements include 
the spatial resolution of the 
diagnostic, the photon statistics of the image, target alignment and 
parallax, and the specific contrast level chosen to define the length of the
Cu spikes. These considerations allowed calculation of the experimental
error bars presented in the figure (described below)
that compares the experimental results to the simulation results.
In addition to the spatial uncertainty, there were also several sources of
uncertainty in the temporal accuracy. These arise
from target-to-target dimensional variations, shot-to-shot drive intensity
variations, and the intrinsic timing accuracy of the diagnostics.
The experimental uncertainty in the timing is, however, relatively small, and is approximately
indicated by the width of the symbols used in the comparison figure (below).

The Flash simulations were two-dimensional with a three-layer arrangement of Cu, 
polyimide CH, and C having the same densities as those of
the experimental target to  model the experiment. 
The initial conditions for the Flash simulations represent the 
configuration 2.1 ns after the laser shot. At this point, the laser
has deposited its energy and the shock is approaching the Cu-CH interface
and the evolution is purely hydrodynamic.
The initial conditions (thermodynamic profiles) for the Flash simulation
were obtained
from simulations of the laser-material interaction performed with a
one-dimensional radiation hydrodynamics code~\citep{larsen94} that
was able to describe the process of energy deposition occurring in
the initial 2.1 ns.
These one-dimensional profiles 
were mapped onto the two-dimensional grid with 
a sinusoidal perturbation added to the Cu-CH interface. 
Figure \ref{fig:shock_init} illustrates the initial configuration of 
the Flash simulations. For convenience, the simulations used periodic 
boundary conditions on the transverse boundaries and zero-gradient 
outflow boundary conditions on the boundaries in the direction of the 
shock propagation. 
The materials were treated as gamma-law gases, with
$\gamma =$ 2.0, 2.0, and 1.3 for the Cu, CH, C,
respectively. These values for gamma were chosen to give
similar shock speeds to the shock speeds observed in the experiments.

From these initial conditions, the simulations
were evolved to
approximately 66 ns. 
\begin{figure}[h]
   \includegraphics[angle=0,
      width=\columnwidth]{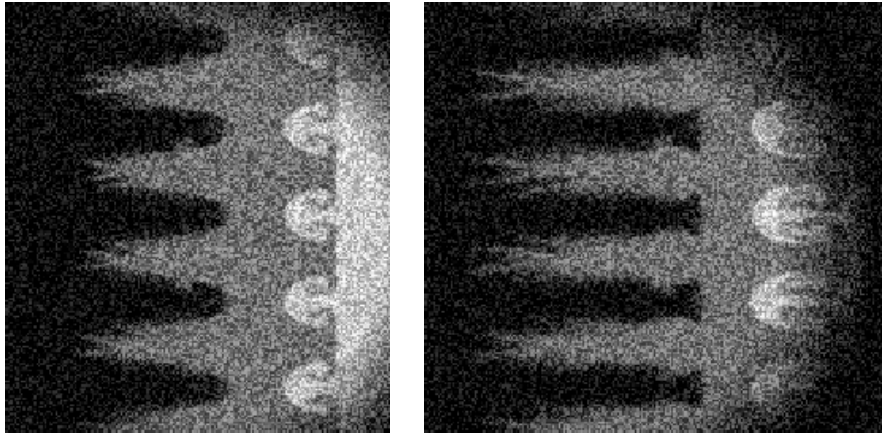}
   \caption{
Simulated radiographs from the 6 levels of refinement (effective resolution of $512 \times
256$) simulation of the three-layer target experiment. The simulated radiographs were 
created from the fluid abundances at times corresponding approximately to those of the 
images from the experiment, 39.9 ns (left) and 66.0 ns (right). Shown are the parts 
of the simulation domain that match the regions in the experimental results.
From \citet{calder.fryxell.ea:on} \copyright\ AAS. Reproduced with permission.
      }
   \label{fig:shock_sim}
\end{figure}
Figure \ref{fig:shock_sim} shows simulated radiographs
from a simulation at an intermediate resolution, allowing 
visual comparison to the
experimental results. The figure presents simulated
radiographs at approximately the two times corresponding to the images from
the experiment, 39.9 ns (left panel) and 66.0 ns (right panel).
The simulation in Figure \ref{fig:shock_sim} had 6 levels of
mesh refinement corresponding to an effective resolution of $1024 \times 512$
grid zones.
The simulated radiographs were created from the abundances of the 
three materials assigning an artificial opacity to each abundance and
applying the opacity to an artificial ``beam." In addition,
the abundances were de-resolved to match the resolution of the pixels
in the experimental images and random Poisson-distributed `noise' was 
added to the intensity.  

An obvious qualitative difference between the simulated and experimental 
radiographs is readily observed in the curvature of the experimental 
instabilities that is not present in the simulations instabilities. 
The use of periodic boundary conditions in
the transverse directions in the simulation was not consistent with the 
boundary conditions of the experiment, which was performed with the three 
materials of the target inside a cylindrical Be shock tube. 
The experiment results
show the influence of the shock tube walls as a curving or pinching of the outer
Cu spikes, while the simulations did not consider these boundary effects.

Comparison of the simulated radiographs to the 
radiographs from the experiment show that the simulations captured the bulk 
behavior of the materials, particularly the growth of Cu spikes and the 
development of C bubbles. We can conclude from this comparison that
the simulations resemble the experimental results. Assessment of
the comparison as``good" or ``bad" is difficult, however, with only a visual 
comparison, especially one that indicates a difference due to a
boundary condition effect.
What is needed is a quantitative comparison,
and for that we apply the same techniques as we apply to verification,
a convergence study to show the simulations converge with resolution
and a quantitative comparison to the experimental results.

To test convergence of the solutions, a suite of simulations 
was performed at increasing resolution.
The effective resolutions of the simulations were
128 $\times$ 64, 256 $\times$ 128, 512 $\times$ 256, 
1024 $\times$ 512,  2048 $\times$ 1024, and 4096 $\times$ 2048,
corresponding
to 4, 5, 6, 7, 8, and 9 levels of adaptive mesh refinement.
As noted above, the lengths of the Cu spikes were chosen as the metric 
for quantitative comparison to the experiments. 
Flash solves an advection equation for each abundance,
which allowed tracking the flow of each material with time.
The spike lengths in the 
simulations were measured by averaging the CH abundance in the 
$y$-direction across the simulation domain then smoothing the resulting
one-dimensional array slightly to minimize differences that would occur
owing to very small scale structure. The length of the Cu spikes was then
determined by the average distance spanned by minimum locations of average 
abundances 0.05 and 0.9. The results were reasonably robust to the amount 
of smoothing and threshold values.

\begin{figure}[h]
   \includegraphics[angle=0,
      width=\columnwidth]{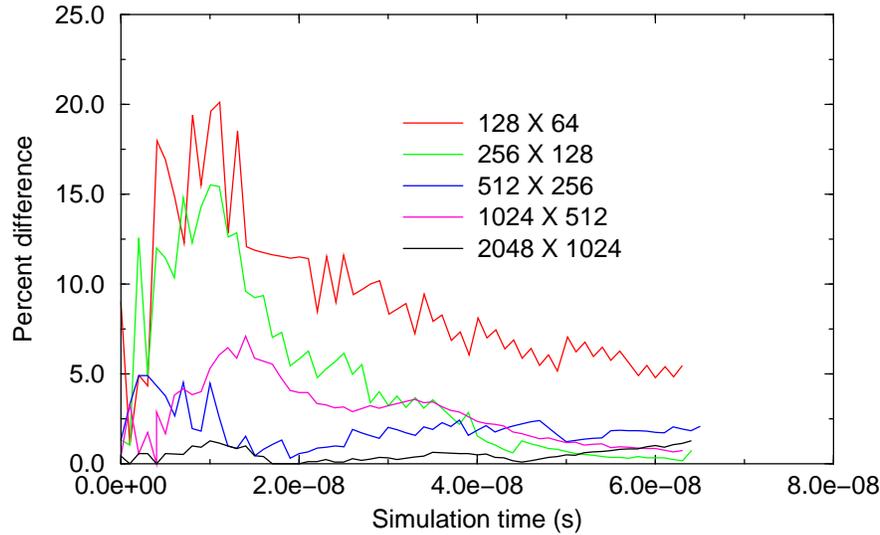}
   \caption{
Percent difference of the Cu spike lengths from those of the
highest resolution (9 levels of adaptive mesh refinement) simulation vs.\
time. The percent differences are from the lower resolution simulations of 4,
5, 6, 7, and 8 levels of adaptive mesh refinement, with the corresponding
effective resolutions in the legend. We note that the convergence 
is not perfect. The curve from the 8 levels of refinement simulation crosses 
those of the 6 and 7 levels of refinement simulations,
indicating a higher percent difference.
Adapted from \citet{calder.fryxell.ea:on}.
      }
   \label{fig:shock_conv}
\end{figure}

The results of testing the convergence of the Cu spike length measurements
are shown in Figure \ref{fig:shock_conv},
which depicts percent differences
from the highest resolution simulation, 9 levels of adaptive mesh
refinement, as functions of time. 
The trend is
that the difference decreases with increasing mesh resolution, with
the 7 and 8 level of adaptive mesh refinement simulations always
demonstrating agreement to within five percent. The trend of decreasing
difference with increasing mesh resolution demonstrates a convergence
of the flow, but it is subject to caveats. We note that the trend does not
describe the behavior at all points in time (that is, the percent difference
curves sometimes cross each other), and this average measurement is an
integral property of the flow and in no way quantifies the differences in
small scale structure observed in the abundances. In particular,
we note that the difference curve for the simulation with 8 levels of 
adaptive mesh refinement 
crosses the curves of both the 7 and 6 level simulations, suggesting
that higher-resolution simulations may deviate further from these results
and produce degraded agreement with the experiment. This result is in keeping
with the above-mentioned concerns with ILES.

\begin{figure}[h]
   \includegraphics[angle=0,
      width=\columnwidth]{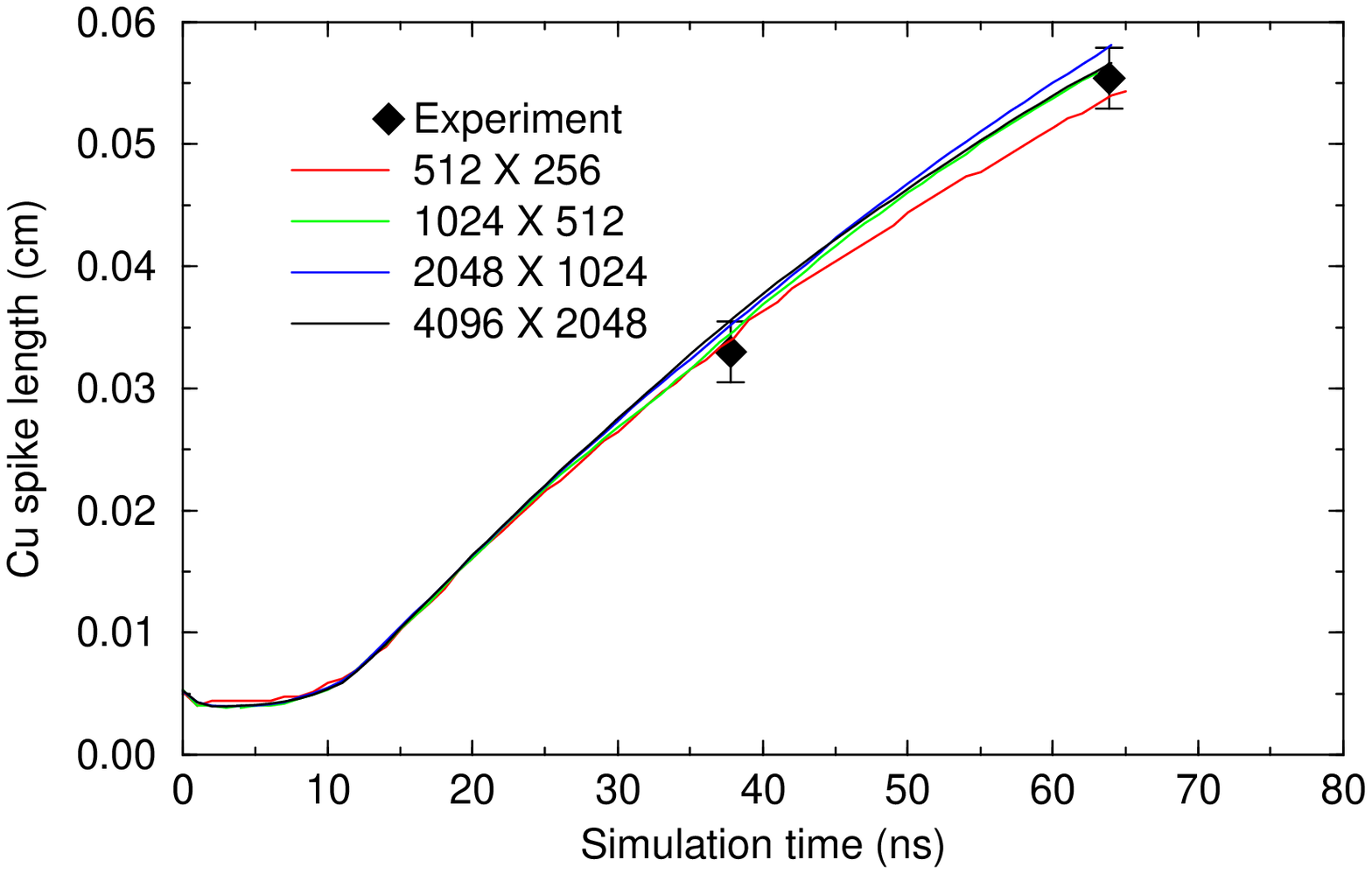}
 \caption{Results from a validation test consisting of a laser-driven
  shock propagating through a multi-layer target. The lengths of the
  Cu spikes is plotted vs.\ time from 4 simulations at 6, 7, 8, and 9 
  levels of adaptive mesh refinement in a convergence study. The effective 
  resolutions are given in the legend. Also shown are the experimental results 
  at two times with spatial error bars of ($\pm 25 \mathrm{\mu m}$). The 
  timing error is about the width of the diamonds marking the
  experimental result. The differences between the simulations at different
  resolutions is less than the uncertainty of the experimental results.
  Adapted from \citet{calder.fryxell.ea:on}.
      }
   \label{fig:shock_results}
\end{figure}

Figure \ref{fig:shock_results} shows the Cu spike length vs.\
time for 4 simulations at increasing resolution. Also shown
are the above-mentioned experimental results. The experimental
error bars correspond to $\pm 25\microm$, the spatial error of the
experiment. The width of the symbols marking the experimental results
indicates approximately the timing error. The figure shows that the
simulations quantitatively agree with the experimental results at the
early and late times to within the experimental uncertainty.

As noted above, this study has previously appeared in the literature. 
Complete details of the validation study may be found in 
\citet{calder.fryxell.ea:on,calder2005,calderetal2006} and additional details of the 
experiments may be found in \citet{calder.pre,robeyetal01}. 

\subsection{Computation of Reaction Products in Large Eddy Simulations of
Supernovae}

When a laboratory experiment is available, the distinction between
verification and validation is fairly clear, as discussed earlier.  However,
when creating predictive simulations of astrophysical processes that cannot
be reproduced directly in the laboratory, even using appropriate scaling
laws, the distinction can become less clear because the task becomes one of 
confirmation of simulation results without laboratory results. In many situations, 
notably in
stellar combustion, it is possible to have a model that is demonstrably
more physically valid but is too expensive or constrained to be used
for the desired predictive simulations. Simpler models must be applied
to simulate observed phenomena, hence the need for comparison of different methods.

Nuclear reaction networks and multi-dimensional simulations present a good
example of this confluence of verification and validation.  In astrophysical
detonations it is possible to compute the steady-state structure of the
propagating reaction front with a large reaction network with hundreds of
species and thousands of reactions using error-controlled numerical methods
\citep[e.g.][]{sharpe1999,mooretownsleybildsten2013}
Consider the following question: How many species are necessary to accurately
predict the characteristics of the flow such as peak temperature and reaction
front width?  This is not a verification question.  We can use verification
techniques to demonstrate that the equations governing the time integration
of the reactions are being solved to a desired accuracy.  Such a test, however, 
does not demonstrate whether or not a particular selection of species is
sufficient for the stated purpose.  So we proceed to compare our model with
say three or a dozen ``effective" reactions or species to another model which we
believe to be more physically valid because it has more complete reaction physics. This situation is 
neither verification that our model is being solved correctly (that is already 
done) nor is it validation against a specific physical experiment.
It is, however, validation under the definition introduced in section \ref{sec:vvapproach} above, in that it addresses whether the model is physically correct.
Some terminology refers to this as confirmation of one model with a physically more valid model.
Since the label depends finely on definitions of terminology, it is useful in discussion to term this type of comparison as something that combines elements of verification and validation (see Ch.\ 41 by Beisbart in this volume).
It is a model-to-model comparison, as verification often is, but addresses the physical applicability of the model, as validation does.

If integration of thousands of reactions were the only issue, this validation
of simplified models might not be worthwhile; instead one would just use the 
better model directly.  There are areas of prediction, however, where direct
use of the better model can be infeasible.  In explosive astrophysical combustion 
(which powers type Ia supernova explosions), it is typically desirable to predict the 
overall products and
the speeds at which they are ejected.  Unfortunately, a simulation that can
predict that information must include the entire star, which may be around
$10^9$~cm in size.  The reaction front through which the combustion takes
place is one cm or less in thickness \citep{townetal15}.  Also, the propagation of this front
through the star will generally occur in a way that obeys no particularly
symmetry, making it necessary to simulate this combustion and ejection of
material in three dimensions.

The necessity of simulating the whole star in three dimensions presents 
several challenges from the
standpoint of V\&V.  First, since the combustion
phenomena occur far below the best possible grid scale ($\sim 10^5$~cm), the
typical method of verification by convergence study is not valid.
Claiming convergence for a numerical solution of differential equations
presupposes that the relevant gradients are numerically resolved and become
better resolved at higher resolution.  This is the very meaning of
resolution.  However, in the full-scale astrophysical case, an
example of the above-mentioned large eddy simulation situation, 
the composition gradients representing the physical reaction
front (the length scale over which the fuel is consumed and converted to
products) are
never actually resolved.
Secondly, while error-controlled methods for ODE integration are
well-understood, similar automated control of accuracy is not available in
current widely used methods for solution of PDEs, such as in hydrodynamics.
Because this control is not built into the method, performing predictive
simulations involves a constant process of verification to ensure that
solutions obtained do not depend on resolution.
That process can be both expensive and time-consuming.
Thirdly, it may be computationally infeasible to
include hundreds of species and thousands of reactions in the full-scale
hydrodynamic simulation, thus even if we were able to verify the methods for
reactive hydrodynamics, we would need to use a model for the reactions that
we know to have specific deficiencies and would therefore need some form of
validation against more physically complete models.
Finally, as discussed earlier, because some physical processes such as fluid
dissipation due to viscosity is left implicit, a higher-resolution simulation
may not only be more numerically accurate but also more physically valid.
As a result of these issues, verification and validation of the simulation of
a stellar explosion can be mixed in a way that is not always cleanly
separable.

Here we will present a discussion of ongoing efforts at verification and
validation of methods for computing the products of thermonuclear supernova
explosions.  The full-star simulations use a simplified model of the
reactions for computational efficiency, and are necessarily under-resolved.
The overall goal is to compare the results from this computational model to
computational models of much higher physical and numerical fidelity.  In the
case of combustion, those are computations with large, complete nuclear
reaction networks computed using resolved, error-controlled numerical
techniques.  The limitation is that the latter methods can only be used under
certain flow conditions, specifically, a steady state.  
We therefore proceed by treating the methods used
in the full-star simulation as the model to be validated by comparison to
more physical calculations.  This is similar to verification by comparison to
a benchmark, except that the two models are known to be different by
construction.

Table \ref{tab:sim_cap} shows a matrix comparing the capabilities of
compressible hydrodynamics simulations in various dimensions as well as the
fully resolved method, which can only be used in one dimension and for
reaction fronts propagating in a steady state through a uniform medium.  As
shown, a resolved calculation with the full network at all densities relevant
to the supernova can only be performed with the steady-state method.
However, this method cannot be used to treat transients (e.g. ignition or
non-spatially uniform density) or general geometries including the full star.
Of the hydrodynamical methods in various spatial dimensions, represented in
the other three columns of the table, only one-dimensional calculations can use
a full reaction network effectively and resolve the reaction front, though
not at all densities.  The possible importance of transient effects
necessitates a multi-step strategy utilizing cross-comparisons of calculations 
of reaction front structure among several different methods.  For example, we 
can verify one-dimensional dynamical calculations at uniform densities using 
comparison to steady-state calculations, and then use one-dimensional 
calculations with non-uniform density to characterize transient
effects.  Even for a transient, it is informative to compare to
steady-state solutions in order to provide physical insight to the importance
of non-uniformities in density.

\begin{table}
\begin{center}

\begin{tabular}{|l|c|c|c|c|}
\hline
Capability
& 3-d & 2-d & 1-d & 1-d steady\\
\hline
\hline
full reaction network
 & $\times$
 & $\times$
 & \cellcolor{green!25} \checkmark
 & \cellcolor{green!25} \checkmark \\
\hline
resolved at low density
 & $\times$
 & $\times$
 & \cellcolor{green!25} \checkmark
 & \cellcolor{green!25} \checkmark \\
\hline
resolved at high density
 & $\times$
 & $\times$
 & $\times$
 & \cellcolor{green!25} \checkmark \\
\hline
transients (dynamical)
 & \cellcolor{green!25} \checkmark
 & \cellcolor{green!25} \checkmark
 & \cellcolor{green!25} \checkmark
 & $\times$
\\\hline
general geometries
 & \cellcolor{green!25} \checkmark
 & $\times$
 & $\times$
 & $\times$
\\\hline
full star
 & \cellcolor{green!25} \checkmark
 & \cellcolor{green!25} \checkmark
 & \cellcolor{green!25} \checkmark
 & $\times$
\\\hline
\end{tabular}

\caption{Capabilities of simulations in various dimensions and assumptions.
Comparison of results among simulations is performed in order to validate
that full star three-dimensional simulations reproduce the results of more
physically valid one- dimensional calculations of steady state properties of
detonations.\label{tab:sim_cap}}
\end{center}

\end{table}

\begin{figure}[h]
   \includegraphics[width=\columnwidth]{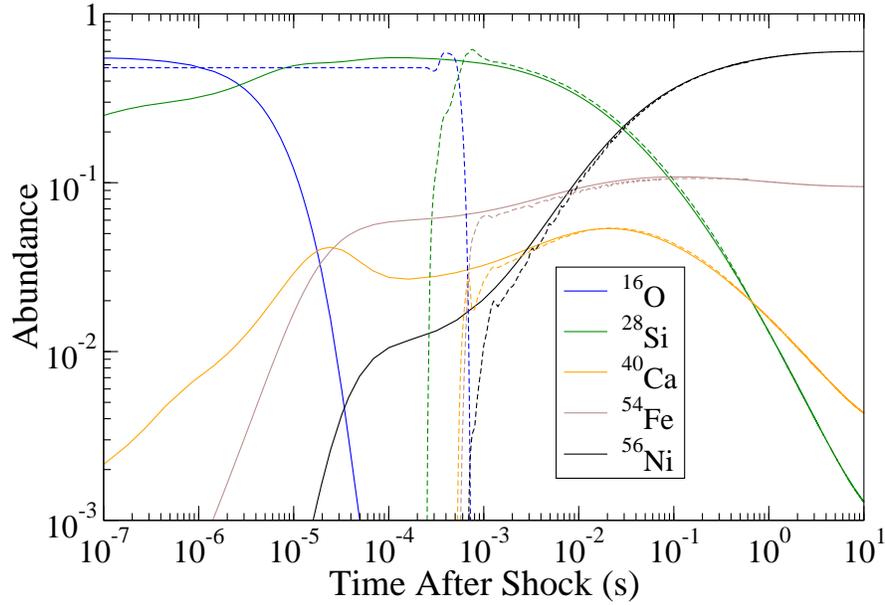}
   \caption{
   Comparison of planar steady-state detonation structure simulated
   hydrodynamically at $10^5$ cm resolution using post-processing of
   Lagrangian tracers (dashed) with the steady-state
   structure computed directly using error-controlled integration (solid).
   Abundances here are given as mass fractions.
   Similar to comparisons made in \citet{townetal15}.
   The oxygen consumption structure will remain unresolvable even with
   more than an order} of magnitude higher resolution in the hydrodynamic
   simulation.
   \label{fig:det_comp}
\end{figure}

Figure \ref{fig:det_comp} shows an example of a comparison of the
compositional structure of a propagating detonation reaction front computed
with the one-dimensional dynamical method and the one-dimensional steady-state
method.  The hydrodynamical simulation (dashed lines) was performed at a
physical resolution of $10^5$~cm, which corresponds to a hydrodynamical
time step of about $10^{-4}$~s.  The fuel here is mostly $^{12}$C and $^{16}$O,
which is reacted to eventually become $^{56}$Ni.  The consumption of $^{12}$C
is not shown, but is even faster than that of $^{16}$O.  The structure for a
detonation propagating in steady state (solid lines) is computed with
an error-controlled method using adaptive time stepping and an error tolerance
of order $10^{-6}$, and is therefore suitably resolved by construction.  The
abundance histories from the hydrodynamical model shown here are the result
of using a simplified reaction model in the hydrodynamics and then
post-processing the resulting density and temperature histories of fluid
elements with a larger reaction network
\citep{travaglioetal2004,townetal15}.  The goal of this comparison is to
validate that away from the unresolved portion of the reaction front
(timescales $\gtrsim 10^{-3}$~s), the composition history is accurately
predicted by the under-resolved calculation with the simplified burning
model.  This comparison shows that the results are in good agreement for
steady-state, planar detonations.  For an example of a comparison for
non-planar (curved) detonations see \cite{mooretownsleybildsten2013}.

The validation of methods for computing astrophysical combustion in large
eddy simulations is ongoing.  The various possible calculations represented
in Table \ref{tab:sim_cap} must be compared for geometries and conditions for
which there is overlap in capability.  This process also entails ongoing improvement
of both the simplified reaction model utilized in the large eddy simulations
\citep{townsleyetal2009,willcoxetal2016} as well as improving techniques for computing the
final yields \citep{townetal15}.

\section{Discussion}

The simulational results for the hydrodynamics validation example fell
within the temporal and spatial error bars of the experimental results
thus showing quantitative  agreement between simulation and experiment
for the metric of the lengths of the copper spikes. This agreement
demonstrates that the hydrodynamics module in Flash captured the bulk
properties of the flow and observable morphology.  and builds confidence
in simulations of astrophysical phenomena. We cannot, however, declare
the code ``validated" for a host of reasons:

\begin{itemize}

\item The experimental configuration produced essentially a two-dimensional
result, hence our modeling it with two-dimensional simulations. The experiment 
was three-dimensional, so correctly describing the fluid instabilities,
particularly the amount of small-scale structure in the flow may 
require three-dimensional simulations. 

\item The models were incomplete. The three materials were modeled as ideal 
gasses, a questionable assumption. Also, for convenience, the simulations 
neglected the presence of the shock tube surrounding the target and assumed 
periodic boundary conditions. Thus the simulations did not 
include effects due to the shock tube.

\item The experimental diagnostics, radiographs, are really shadows that
cannot adequately capture small-scale structure. Even if three-dimensional
simulations that better described the fluid instabilities had been performed,
comparison to the experimental results is limited by the experimental diagnostics.

\item The observed degraded agreement between simulations at the highest 
resolutions indicates the results are not converged. We attribute this
result to the fact that the Euler equations allow a changing numerical viscosity
with resolution, which changes the Reynolds number and thus the nature of any
turbulence. Additional commentary on this issue may be found in \citet{calder.fryxell.ea:on}.

\end{itemize}

Even with limitations, the demonstrated ability
of the simulations to capture the expected bulk properties of the
flow builds confidence in the results of astrophysical simulations,
allowing us to conclude that the shocks and fluid instabilities study 
was a success. The principal
differences observed between the results from simulations and the 
experimental results were in the amount of small scale structure observed
in the flow, with the amount of small-scale structure in the
simulations increasing with resolution. This behavior is expected
because the effective Reynolds number increases with resolution
as described above, and we believe this effect is the source of the 
observed imperfect convergence. 
Because the experimental data are
radiographs and cannot capture the actual amount of small-scale 
structure in the flow, the correct amount of small-scale structure
remains undetermined and even if the convergence of the simulations
had been perfect, we could not conclude the solution converged
to the correct result.

In addition to increasing confidence in the results, the hydrodynamics 
validation study was well worth the investment because of the lessons 
learned in comparing the experimental and simulational results. The 
collaborative process of determining the metric for comparison and 
extracting the results from the experimental and simulational data 
resulted in a better understanding of the issues, which also 
increases confidence in the astrophysical results. 
The experimentalists also benefited from the process of validation
because the process of comparison suggested metrics for future comparisons,
provided useful diagnostics, and supplied a virtual model that
aided in the design of future experiments. A point worth stressing again 
in conclusion is the importance of close collaboration between the 
experimentalists and theorists needed to make a meaningful quantitative 
comparison.  Raw experimental data such as a radiograph alone does not 
allow for
a quantitative comparison to simulational results. Finally, we note that
the success of this collaboration seeded interest in high energy density
physics among the developers of Flash, which subsequently resulted in
an extended course of collaborative research into high energy density 
physics (see \citet{Tzeferacosetal2015} and references therein).

The product of reactive hydrodynamics study gave a look at the process
of comparing models of differing fidelity to ensure that macroscopic
(three-dimensional) simulations capture the physics of thermonuclear
reactions while also allowing the calculation of detailed abundances. 
Our approach is to test simplified models against higher-fidelity models 
for a given physical process, here thermonuclear combustion. Simplified 
models then facilitate three-dimensional 
simulations that would be intractable otherwise. The 
results of these studies are also applicable to the problem of determining 
detailed abundances from the density and temperature histories of Lagrangian 
tracers. We illustrated this process with a comparison between results from 
post-processed tracers from a hydrodynamics simulation and a detailed 
calculation of steady-state burning structure. This study confirmed that
our simulations capture the essence of the reactions in whole-star models,
and thereby increased confidence in our predictions of the astrophysical 
events.

\section{Conclusions}

The cases we present here are but one part of the continuing effort at
verifying and validating Flash and associated infrastructure (e.g.\ 
the post-processing method presented here). The first study of
validating the hydrodynamics was performed early in the 
development of Flash. Though very informative, it could
have been continued further with additional quantification of 
the effect of missing physics as a good next step. Also,
further modifications to the code would allow it to capture 
high energy density phenomena better. Such activities, however,
were not critical to the astrophysical problems.  Still, the
case was very informative and served to increase confidence
in the results. The second case, the computation of reaction
products in large eddy simulations of supernovae, is very much
a work in progress and represents our contemporary effort.

Our conclusion from both of these studies is that like any
discipline in computational science, V\&V are an essential part of
the process of modeling astrophysical phenomena. V\&V in astrophysics 
can be particularly challenging due to the inaccessibility of the 
physical conditions attained and limited ancillary measurements available 
for distant events.  As shown here by these examples, however, positive 
steps that build confidence in models can be taken based on comparisons 
using related laboratory experiments and more complete physical models 
where available.

\begin{acknowledgement}
This contribution was supported in part by the Department of Energy through
grant DE-FG02-87ER40317 and the research described here was supported in part
by earlier grant B341495 to the Center for Astrophysical
Flashes at the University of Chicago. The software used in this work was in part 
developed by the DOE-supported ASC/Alliances Center for Astrophysical Thermonuclear 
Flashes at the University of Chicago. This research has made use of NASA's Astrophysics 
Data System. The authors thank Eric Winsberg and Bruce Fryxell for helpful discussions about this
manuscript.
\end{acknowledgement}

\newcommand{\apj}{Astrophysical Journal}
\newcommand{\mnras}{Monthly Notices of the Royal Astronomical Society}
\newcommand{\aap}{Astronomy and Astrophysics}

\end{document}